\documentclass[useAMS,usenatbib]{mn2e}

\usepackage{epsfig}
\usepackage{multirow}
\usepackage{color}

\def\gsim{\;\rlap{\lower 2.5pt
 \hbox{$\sim$}}\raise 1.5pt\hbox{$>$}\;}
\def\lsim{\;\rlap{\lower 2.5pt
   \hbox{$\sim$}}\raise 1.5pt\hbox{$<$}\;}
\def\ie{{\it i.e. }}
\def\eg{{\it e.g. }}

\def\vti{\vec{\theta}^I}
\def\vts{\vec{\theta}^S}

\def\vt{\vec{\theta}}

\title[Simple Treatments of the Photon Noise and the Pixelation Effect in Weak Lensing]{Simple Treatments of the Photon Noise and the Pixelation Effect in Weak Lensing}

\author[Jun Zhang]
{Jun Zhang\thanks{E-mail:jzhang@astro.as.utexas.edu}\\ 
\\
Department of Astronomy, University of California, Berkeley, CA 94720, USA\\
Texas Cosmology Center, the University of Texas at Austin, Austin, TX, 78712, USA\\ 
}

\begin{document}

%\date{Accepted . Received ; in original form }    

\pagerange{\pageref{firstpage}--\pageref{lastpage}} \pubyear{2006}

\maketitle

\label{firstpage}
                                      
\begin{abstract}

We propose easy ways of correcting for the systematic errors caused by the photon noise and the pixelation effect in cosmic shear measurements. Our treatment of noise can reliably remove the noise contamination to the cosmic shear even when the flux density of the noise is comparable with those of the sources. For pixelated images, we find that one can accurately reconstruct their corresponding continuous images by interpolating the logarithms of the pixel readouts with either the Bicubic or the Bicubic Spline method as long as the pixel size is about less than the scale size of the point spread function (PSF, including the pixel response function), a condition which is almost always satisfied in practice. Our methodology is well defined regardless of the morphologies of the galaxies and the PSF. Despite that our discussion is based on the shear measurement method of Zhang (2008), our way of treating the noise can in principle be considered in other methods, and the interpolation method that we introduce for reconstructing continuous images from pixelated ones is generally useful for digital image processing of all purposes.   

\end{abstract}


\section{Introduction}
\label{intro}

Weak gravitational lensing has been widely used as a direct probe of the large scale structure (see reviews by \citealt{bs01,wittman02,refregier03}). By measuring the systematic distortions of background galaxy images, one can place constraints on the cosmological parameters (\citealt{bre00,kwl00,vw00,wittman00,maoli01,rhodes01,vw01,hyg02,rrg02,bmre03,brown03,hamana03,jarvis03,rhodes04,heymans05,mbre05,vmh05,dahle06,hoekstra06,jjbd06,semboloni06,hetterscheidt07,schrabback07}). With accurate redshift information, the geometry and the structure growth rate of our Universe can be constrained as functions of redshift separately, providing a consistency test of the gravity theory (\citealt{hu02,abazajian03,jain03,acquaviva04,bernstein04,hu04,kratochvil04,song04,takada04,takadawhite04,ishak05,simpson05,song05,zhang05,hannestad06,ishak06,zhan06,knox06,schimd07,taylor07}). 

A key issue in weak lensing is about how to measure the cosmic shear with galaxy shapes. This is difficult mainly because the signal-to-noise ratio of the measurement on one galaxy is typically only a few percent. It is therefore extremely important for any shear measurement method to carefully treat any possible systematic errors, including at least the following: the correction due to the image smearing by the point spread function (PSF, including the pixel response function); the photon noise; the pixelation effect due to the discrete nature of the CCD pixels. There have been a number of methods proposed to deal with the corrections due to the PSF (see \citealt{tyson90,bonnet95,kaiser95,luppino97,hoekstra98,rhodes00,kaiser00,bridle01,bernstein02,refregierbacon03,massey05,kuijken06,miller07,nakajima07,kitching08,zhang08}). However, the photon noise and the pixelation effect remain to be treated in a more systematic way. For example, existing model fitting methods (\eg, \citealt{bridle01,bernstein02,kuijken06,miller07,nakajima07,kitching08}) use the variance of the noise to weight the pixels in their chi-square fittings. It is not clear to what level the noise contamination to the shear recovery can be removed in this way, especially for correlated background noise (see, \eg, \citealt{massey07}). By integrating the model over the pixels, the model fitting methods essentially use linear interpolations to treat the pixelation effect. This is found not accurate as we will discuss in \S\ref{pixelation}. Indeed, as we will discuss later in the paper, there are two types of noise: the astronomical photon noise and the "photon counting" shot noise. While the second type diminishes when the exposure time increases, the first type does not. We will mainly focus on the astronomical noise in this paper. For the counting shot noise, we will simply argue in \S\ref{summary} that its contamination to the shear recovery can be significantly suppressed by increasing the exposure time. Since these issues are not specifically addressed in any previous weak lensing literatures, it is important to point them out in this paper. 

In a recent work by Zhang (2008) (Z08 hereafter), a new and simple way of measuring the cosmic shear is found. Its main advantages includes: 1. it is mathematically simple; 2. it is free of assumptions on the morphologies of the galaxies and the PSF; 3. it enables us to probe the shear information from galaxy substructures, thereby improving the signal-to-noise ratio. These facts encourage us to extend the method further by including a treatment of the photon noise and the pixelation effect. Fortunately, we find that these two types of systematic errors can be treated in a simple and model-independent way based on the method of Z08. As will become clear later in this paper, the method we adopt to remove the noise contamination can also be considered in other shear measurement methods, and our treatment of the pixelation effect is generally useful for image processing of all purposes.

The paper is organized as follows: in \S\ref{review}, we briefly review the shear measurement method of Z08; in \S\ref{systematics}, we show how to treat the photon noise (in \S\ref{noise}) and the pixelation effect (in \S\ref{pixelation}) in weak lensing; in \S\ref{numerics}, we use computer-generated mock galaxy images to test the performance of our method; finally, we summarize and discuss remaining issues in \S\ref{summary}.

%-------------------------------------------------------------------
\section{Review of the Shear Measurement Method}
\label{review}

Z08 proposes a way of measuring the cosmic shear with the spatial derivatives of the galaxy surface brightness field. To do so, let us define the surface brightness on the image plane as $f_I(\vti)$, and that on the source plane as $f_S(\vts)$, where $\vti$ and $\vts$ are the position angles on the image and source plane respectively. These quantities are related in a simple way as:
\begin{eqnarray}
\label{fifstits} 
&&f_I(\vti)=f_S(\vts)\\ \nonumber
&&\vti=\mathbf{A}\vts
\end{eqnarray}
where $\mathbf{A}_{ij}=\delta_{ij}+\Phi_{ij}$, and $\Phi_{ij}=\partial\delta\theta^I_i/\partial\theta^S_j$ are the spatial derivatives of the lensing deflection angle. Matrix $\mathbf{A}$ is often expressed in terms of the convergence $\kappa=(\Phi_{xx}+\Phi_{yy})/2$ and the two shear components $\gamma_1=(\Phi_{xx}-\Phi_{yy})/2$ and $\gamma_2=\Phi_{xy}$. Assuming the intrinsic galaxy image $f_S(\vts)$ is statistically isotropic, the shear components can be simply related to the derivatives of the surface brightness field as (\citealt{seljak99}):
\begin{eqnarray}   
\label{shear12}
\gamma_1&=&-\frac{1}{2}\frac{\langle (\partial_xf_I)^2-(\partial_yf_I)^2\rangle}{\langle (\partial_xf_I)^2+(\partial_yf_I)^2\rangle} \\ \nonumber
\gamma_2&=&-\frac{\langle\partial_xf_I\partial_yf_I\rangle}{\langle (\partial_xf_I)^2+(\partial_yf_I)^2\rangle}
\end{eqnarray}
where the averages are taken over the galaxy.

Eq.[\ref{shear12}] is useful only when the angular resolution of the observation is infinitely high. In practice, the observed galaxy surface brightness distribution $f_O$ is always equal to the lensed galaxy image $f_I$ convoluted with the PSF, \ie:
\begin{equation}
\label{fofi}
f_O(\vt)=\int d^2\vti W(\vt -\vti)f_I(\vti)
\end{equation}  
where $W$ is the PSF. Z08 has shown how to modify eq.(\ref{shear12}) when the PSF is an isotropic Gaussian function, which can be written as:
\begin{equation}
\label{wbeta}
W(\vt)=\frac{1}{2\pi\beta^2}\exp\left(-\frac{\vert\vt\vert^2}{2\beta^2}\right)
\end{equation}
where $\beta$ is the scale radius of the Gaussian function. The new relation between the shear components and the derivatives of the surface brightness field is:
\begin{eqnarray}
\label{shear12PSF}
\gamma_1&=&-\frac{1}{2}\frac{\langle (\partial_xf_O)^2-(\partial_yf_O)^2\rangle}{\langle (\partial_xf_O)^2+(\partial_yf_O)^2\rangle+\Delta} \\ \nonumber
\gamma_2&=&-\frac{\langle\partial_xf_O\partial_yf_O\rangle}{\langle (\partial_xf_O)^2+(\partial_yf_O)^2\rangle+\Delta}
\end{eqnarray}
where
\begin{equation}
\label{Delta}
\Delta=\frac{\beta^2}{2}\langle\vec{\nabla}f_O\cdot\vec{\nabla}(\nabla^2f_O)\rangle
\end{equation}

For a general PSF, one can transform it into the desired isotropic Gaussian form through a convolution in Fourier space. The scale radius $\beta$ of the target PSF should be larger than that of the original PSF to avoid singularities in the convolution. Furthermore, as shown in Z08, the spatial derivatives required by eq.(\ref{shear12PSF}) can also be easily evaluated in Fourier space.

%----------------------------------------------------------------------
\section{Treating the Photon Noise and the Pixelation Effect}
\label{systematics}

In this section, we introduce the basic ideas for treating the photon noise and the pixelation effect in \S\ref{noise} and \S\ref{pixelation} respectively. Numerical examples are given in the next section. 

\subsection{Photon Noise}
\label{noise}

First of all, there are two types of photon noise: the astronomical photon noise due to the fluctuation of the background, and the "photon counting" shot noise due to finite exposure time. Note that the first type of noise, like the source, is convoluted by the PSF, while the second type of noise varies from pixel to pixel even if the pixel size is much smaller than the PSF size. In the rest of the paper, we mainly deal with the astronomical noise. For the "photon counting" shot noise, we will simply argue in \S\ref{summary} that its contamination to the shear recovery can be significantly suppressed by increasing the exposure time. 

The presence of the photon noise makes the measurement of the cosmic shear more complicated in two ways: 1. the observed surface brightness $f_O$ is from both the lensed source and the un-lensed foreground noise; 2. because of the aliasing power caused by the non-periodic boundaries of the {\it noisy} map, the measurement of the spatial derivatives of the surface brightness field cannot be accurately performed in Fourier space. Note that simple treatments such as filtering out the noise outside of the source image do not completely fix this problem, because the noise inside the image can still bias the shear estimate. Fortunately, as we show in the rest of this section, our master equation [eq.(\ref{shear12PSF})] for estimating the cosmic shear can be easily adapted to solve both problems.

In the method of Z08, to isolate the source signals in a noisy map, let us first write the total observed surface brightness $f_O$ as the sum of the contributions from the source $f^s$ and the noise $f^n$, \ie, $f_O=f^s+f^n$. Note that in this case, instead of $f_O$, $f^s$ should be used in eq.(\ref{shear12PSF}) to correctly measure the shear components. For this purpose, let us use the following relation:
\begin{eqnarray}
\label{OSN}
\langle (\partial_xf^s)^2\rangle&=&\langle (\partial_xf_O)^2\rangle-\langle (\partial_xf^n)^2\rangle\\ \nonumber
&-&2\langle \partial_xf^s\partial_xf^n\rangle\\ \nonumber
\langle (\partial_yf^s)^2\rangle&=&\langle (\partial_yf_O)^2\rangle-\langle (\partial_yf^n)^2\rangle\\ \nonumber
&-&2\langle \partial_yf^s\partial_yf^n\rangle\\ \nonumber
\langle \partial_xf^s\partial_yf^s\rangle&=&\langle\partial_xf_O\partial_yf_O\rangle-\langle \partial_xf^n\partial_yf^n\rangle\\ \nonumber
&-&\langle \partial_yf^s\partial_xf^n\rangle-\langle \partial_xf^s\partial_yf^n\rangle\\ \nonumber
\langle\vec{\nabla}f^s\cdot\vec{\nabla}(\nabla^2f^s)\rangle&=&\langle\vec{\nabla}f_O\cdot\vec{\nabla}(\nabla^2f_O)\rangle-\langle\vec{\nabla}f^n\cdot\vec{\nabla}(\nabla^2f^n)\rangle\\ \nonumber
&-&\langle\vec{\nabla}f^s\cdot\vec{\nabla}(\nabla^2f^n)\rangle-\langle\vec{\nabla}f^n\cdot\vec{\nabla}(\nabla^2f^s)\rangle
\end{eqnarray}
For simplicity, in this paper, we only consider the photon noise that is from the foreground or the instruments. The surface brightness distribution of the noise is therefore uncorrelated with that of the background sources\footnote{More general cases (\eg, photon noise coming from faint background sources) are more complicated, and will be dealt with in a future work.}. Under this assumption, the cross-correlations between the source and the noise terms (such as, \eg, $\langle \partial_xf^s\partial_xf^n\rangle$) should vanish. Eq.(\ref{OSN}) therefore becomes:
\begin{eqnarray}
\label{OSN3}
\langle (\partial_xf^s)^2\rangle&=&\langle (\partial_xf_O)^2\rangle-\langle (\partial_xf^n)^2\rangle\\ \nonumber
\langle (\partial_yf^s)^2\rangle&=&\langle (\partial_yf_O)^2\rangle-\langle (\partial_yf^n)^2\rangle\\ \nonumber
\langle \partial_xf^s\partial_yf^s\rangle&=&\langle\partial_xf_O\partial_yf_O\rangle-\langle \partial_xf^n\partial_yf^n\rangle\\ \nonumber
\langle\vec{\nabla}f^s\cdot\vec{\nabla}(\nabla^2f^s)\rangle&=&\langle\vec{\nabla}f_O\cdot\vec{\nabla}(\nabla^2f_O)\rangle-\langle\vec{\nabla}f^n\cdot\vec{\nabla}(\nabla^2f^n)\rangle
\end{eqnarray}
The relations in eq.(\ref{OSN3}) suggest an easy way of removing the contaminations from the photon noise: one can use a neighboring map of pure noise $f^n$ to estimate $\langle (\partial_xf^n)^2\rangle$, $\langle (\partial_yf^n)^2\rangle$, $\langle \partial_xf^n\partial_yf^n\rangle$, $\langle\vec{\nabla}f^n\cdot\vec{\nabla}(\nabla^2f^n)\rangle$, and subtract them from their counterparts evaluated from the noisy source map $f_O$ to get the source terms required by eq.(\ref{shear12PSF}). Note that since the noise photons are distributed differently in each map, the above procedure does not exactly remove the noise contribution for each source image. However, the method is statistically accurate as long as the statistical properties of the photon noise are stable over a reasonably large scale. In other words, the differences in the noise distributions of two maps add statistical errors to the measured cosmic shear through this procedure, but no systematic errors. Finally, the pure noise map should be a close neighbor of the source map so that they share the same point spread function.  

To evaluate the derivatives of the surface brightness field of a noisy map in the Fourier space, we need to deal with the non-periodic boundaries appropriately to avoid aliasing powers. This can be done by gradually attenuating the noise towards the boundaries of the map. The attenuation can take an arbitrary form as long as the following criterions are satisfied: 1. the source region is not affected; 2. the edges of the map should be rendered sufficiently faint; 3. the attenuation amplitude should not have abrupt spatial variations; 4. to properly remove the noise contamination, the same attenuation should also be applied to the neighboring map of pure noise. 

In \S\ref{numerics}, we show numerical examples to support the noise treatment discussed above.

\subsection{The Pixelation Effect}
\label{pixelation}

Modern astronomical images are commonly recorded on CCD pixels, the discrete nature of which may affect the accuracy of the cosmic shear measurements. In general, to avoid significant shear measurement errors, the pixel size should be at least a few times smaller than the scale radius (or FWHM) of the PSF. For instance, we find that the method of Z08 requires the scale radius of the PSF to be roughly 3 or 4 times larger than the pixel size. This requirement is often not satisfied in space-based observations. It is therefore useful to have a method which can reconstruct continuous images from under-sampled ones. 

This is indeed a well-defined interpolation problem: how to reconstruct a continuous function if its value is only given at a set of discrete points. In the context of weak lensing, there is a quantitative way of testing the performance of the interpolation method, which is to check the accuracy of the shear recovery using the reconstructed images. The best method should yield the fastest convergence to an accurate image reconstruction (or shear recovery) as the pixel size becomes smaller. 

We have tested several standard 2D interpolation methods, including Bilinear interpolation, Bicubic interpolation, and Bicubic Spline interpolation (see \citealt{press92} for details). Although these conventional methods perform reasonably well, we find that they can all be significantly improved by interpolating the natural logarithms of the data instead of the data themselves. This is mainly due to two reasons: 1. the values of the data have a lower bound --- zero; 2. at large distances, the PSF typically falls off exponentially. For convenience, in the rest of the paper, we call such extensions of the three classical interpolation methods their original names with the prefix ``Log-'', and always abbreviate ``Bicubic Spline'' to ``Spline''. The mathematical definitions of the six methods (\ie, Bilinear, Bicubic, Spline, Log-Bilinear, Log-Bicubic, Log-Spline) are given in the appendix. In \S\ref{numerics}, we show that the Log-Bicubic and Log-Spline methods are most accurate among the six. As will be shown in our numerical examples, continuous images that are reconstructed by these two interpolation methods yield negligible systematic errors in shear recovery as long as the pixel size is about smaller than the PSF size (twice its scale radius). Note that the pixel size is {\it rarely} larger than the PSF size in practice, because the pixel response function is a part of the PSF.

\section{Numerical Examples}
\label{numerics}

We present numerical examples to support the ideas introduced in the previous section. The general setup of our numerical simulations are given in \S\ref{setup}. In \S\ref{test_noise} and \S\ref{test_pixelation}, we test our treatments of the photon noise and the pixelation effect separately. Finally, the overall performance of our method is shown in \S\ref{overall}.

\subsection{General Setup}
\label{setup}

Each of our galaxy images is placed on a $2^n\times 2^n$ grid, where $n$ is an integer (typical chosen to be $8$). Note that such a choice facilitates the Fast Fourier Transformation (FFT). Each grid point is treated as the location of the center of a CCD pixel, whose side length is equal to the grid size. All sizes in our simulations are expressed in units of the grid size, \ie, the pixel size. 

\begin{figure}
\centering
\includegraphics{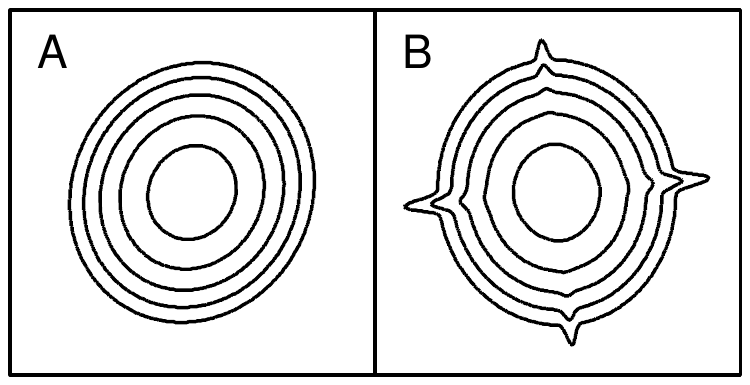}
\caption{Two PSF forms used in this paper. The letter on the up-left corner of each plot is the label of the PSF defined in eq.(\ref{Ws}) (rotated by certain angles). The contours mark 0.0025\%, 0.025\%, 0.25\%, 2.5\%, and 25\% of the peak intensity.  
}
\label{PSFs}
\end{figure}

The form of the PSF is chosen from the following two functions rotated by certain angles:
\begin{eqnarray}
\label{Ws}
&&W_A(x,y)\propto \exp\left[-\frac{1}{2R_{PSF}^2}\left(x^2+0.8y^2\right)\right]\\ \nonumber
&&W_B(x,y)\propto \exp\left[-\frac{1}{2R_{PSF}^2}\left(x^2+0.8y^2\right)\right]\\ \nonumber
&+&0.03\exp\left[-\left(\frac{x^2}{R_{PSF}^2}+0.2\right)\cdot\left(\frac{y^2}{R_{PSF}^2}+0.2\right)\right]
\end{eqnarray}
where $R_{PSF}$ is the scale radius of the PSF. A schematic view of the PSF functions is shown in fig.\ref{PSFs}. Note that the additional term in $W_B$ mimics the diffraction spikes. As discussed in \S\ref{review}, before measuring the shear using eq.(\ref{shear12PSF}), the PSF is always transformed into the desired isotropic Gaussian form, whose scale radius $\beta$ should be slightly larger than $R_{PSF}$ defined here to avoid singularities in the transformation. Note that we have reserved the Greek letter $\beta$ for the scale radius of the target PSF to distinguish it from $R_{PSF}$. 

Both the galaxies and the noise are treated as collections of point sources. For example, the image of a galaxy in our simulation is typically made of a few hundred or thousand points. The advantage of doing so is that one can easily lense the galaxy by displacing its point sources and modifying their amplitudes. The intensities of the point sources are distributed to the neighboring grid points of their locations according to the PSF. For example, a point source of intensity $A$ at location $\vec{x}$ contributes an intensity of $A\cdot W_{PSF}(\vec{y}-\vec{x})$ to the grid point at location $\vec{y}$.  The total intensity on a grid point is the sum of contributions from all the point sources. Since everything is composed of point sources in our simulation, we will mostly call the surface brightness the flux density in the rest of the paper.

There are two types of mock galaxies we use in this paper: regular disk galaxies and irregular galaxies. Our regular galaxy contains a thin circular disk of an exponential profile and a co-axial de Vaucouleurs-type bulge (\citealt{vaucouleurs91}). On average, its face-on surface brightness distribution is parameterized as:
\begin{equation}
\label{dgalaxy}
f(r)\propto\exp(-r/r_{disk})+f_{b/d}\exp\left[-\left(r/r_{bulge}\right)^{1/4}\right]
\end{equation}
where $r$ is the distance to the galaxy center, $r_{bulge}$ and $r_{disk}$ are the scale radii of the bulge and the disk respectively, and $f_{b/d}$ determines the relative brightness of the bulge with respect to the disk. In the simulation, this profile is realized by properly and randomly placing a certain number (typically a few hundred) of point sources. These point sources are projected onto a randomly oriented image plane, lensed, and finally assigned to the CCD pixels according to the PSF to yield the galaxy image. Our irregular galaxies are generated by 2D random walks. The random walk starts from the center of the grid, and continues for a certain number of steps. Each step has a fixed size and a completely random orientation in the image plane. The joint of every two adjacent steps gives the pre-lensing position of a point source of the galaxy in the image plane. The resulting irregular galaxies usually contain abundant substructures. 

For numerical manageability, we always cutoff the galaxy profile at a certain radius $R_G$, which is denoted as the scale radius of the galaxy. This is done by excluding the points that are outside of radius $R_G$ in generating our regular galaxies. For the irregular galaxies, the random walker is sent back to the origin to continue from there when it reaches the radius $R_G$. 

Without loss of generality, we set $f_{b/d}=1/3$, $r_{disk}/r_{bulge}=3$, and $r_{disk}=R_G$ for the regular galaxies.

\subsection{Testing the Photon Noise Treatment}
\label{test_noise}

\begin{figure}
\centering
\includegraphics{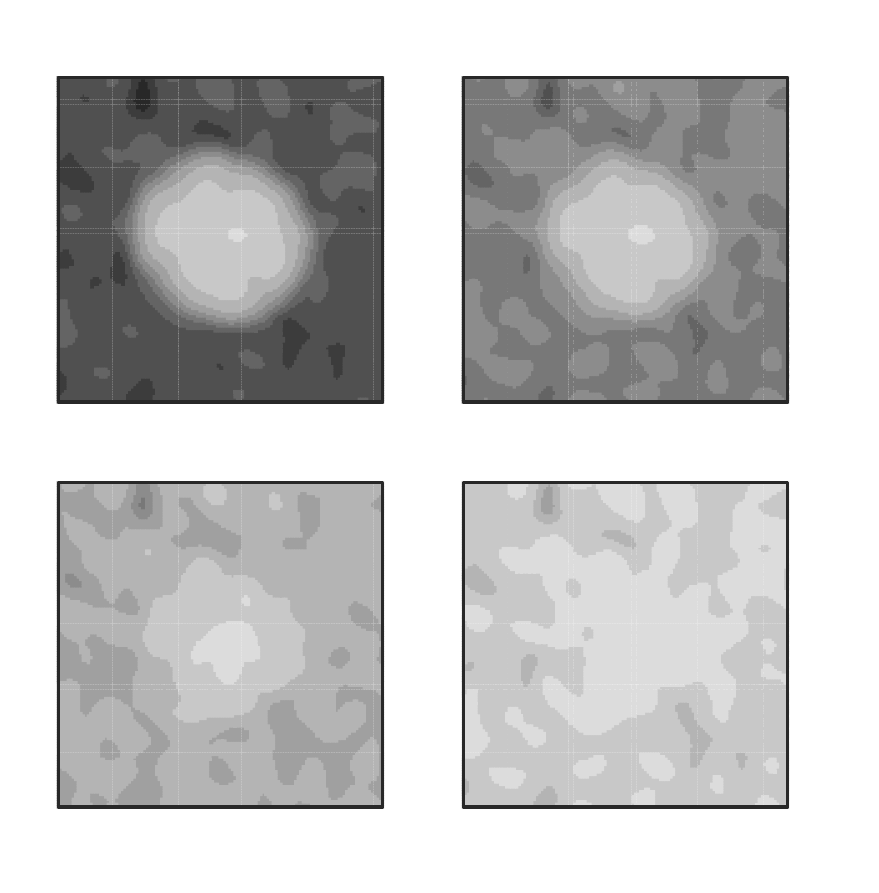}
\caption{Sample images of four different noise-to-galaxy flux density ratios. The up-left, up-right, lower-left, lower-right plots are for flux density ratios of $0.02$, $0.1$, $0.5$, $1.5$ respectively.
}
\label{galaxymap_noise}
\end{figure}

As our first example, we use $10000$ mock regular galaxies to test the treatment of photon noise discussed in \S\ref{noise}. Each galaxy is made of $600$ point sources. The galaxy size $R_G$ is fixed at $30$. The angle between the line-of-sight direction and the normal vector of the disk plane is randomly chosen from $[0, \pi /6]$. The PSF we use is $W_A$ of eq.(\ref{Ws}) with $R_{PSF}=4$ (no pixelation effect). The scale radius $\beta$ of the target isotropic Gaussian PSF is $6$. The noise is made of Poisson distributed point sources, the mean number density of which is roughly one per $4\times 4$ area. Each point source of the noise is given the same intensity, the value of which can be freely adjusted to determine the ratio of the mean flux density of the noise to that of the galaxy (within its scale radius). Note that a separate pure noise map is generated for each noisy galaxy map for removing the noise contaminations to the shear signal. Fig.\ref{galaxymap_noise} shows sample images of four different noise-to-galaxy flux density ratios: $0.02$, $0.1$, $0.5$, $1.5$. To avoid aliasing powers in Fourier transformation, the noise near the boundaries of the map is filtered out by a window function defined as follows:
\begin{eqnarray}
\label{window_noise}
W_{filter}(r)= \left \{ \begin{array}{ll}
1 & \mbox{if $r\le R_{core}$};\\ \nonumber
\exp\left[-\frac{1}{2}\left(\frac{r-R_{core}}{R_{width}}\right)^2\right] & \mbox{if $r > R_{core}$}.\end{array} \right.
\end{eqnarray}
where $r$ is the distance to the map/galaxy center, $R_{core}$ is the radius encircling the unaffected region, and $R_{width}$ determines the width of the transition area. Note that the flat core of the filter should be sufficiently large to avoid affecting the galaxy images. In our simulations, we always choose $R_{core}=R_G+4\beta$, and $R_{width}=\beta$. Finally, to correct for the noise, for each galaxy map, we generate a map of pure noise to measure the noise properties required by eq.(\ref{OSN3}). The same filter is also applied to the pure noise map before the Fourier transformation. 

\begin{figure}
\centering
\includegraphics{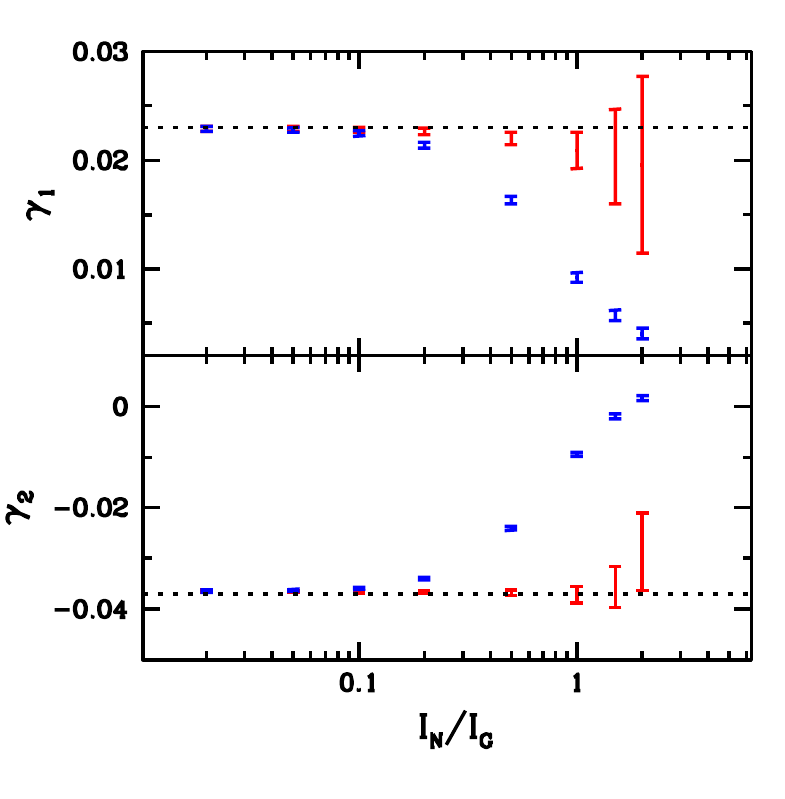}
\caption{The shear values with $1\sigma$ error bars measured from images of different noise-to-galaxy flux density ratios $I_N/I_G$. The measurement uses $10000$ mock regular galaxies. The dotted lines indicate the input shear values. The red data points are our main results achieved through the complete treatment of noise introduced in \S\ref{noise}. The blue ones show the shear values measured directly from the noisy galaxy maps using eq.(\ref{shear12PSF}) without removing the noise contaminations. 
}
\label{shear_regular_noise_only}
\end{figure}

In fig.\ref{shear_regular_noise_only}, we plot the recovered shear values for different noise-to-galaxy flux density ratios $I_N/I_G$. The input cosmic shear $(\gamma_1, \gamma_2)$ is $(0.023, -0.037)$, displayed as dotted lines. The red data points with $1\sigma$ error bars are our main results achieved through the complete treatment of noise. For a comparison, the blue ones show the shear values measured directly from the noisy galaxy maps using eq.(\ref{shear12PSF}) without correcting for the noise (but the filter given by eq.(\ref{window_noise}) is still applied to avoid the aliasing power in Fourier transformation). The figure clearly demonstrates that our noise treatment works remarkably well even when the mean flux density of the noise is comparable with that of the galaxy, though we caution that the size of the error bar grows with increasing noise intensity. On the other hand, the blue data points indicates that without a proper treatment of the noise, the measured shear values quickly drops to zero as the noise flux becomes dominant, deviating significantly from the input shear values.  

Finally, it is worth noting that both here and in the simulations reported in the rest of the paper, the number density of the noise points is always chosen to be roughly one point per PSF area ($\sim R_{PSF}^2$) or slightly more. This is due to two reasons: 1. for a fixed mean noise flux density, a higher number density of the Poisson distributed noise points indeed leads to lower spatial fluctuations of the noise surface brightness field, therefore a less contamination to the shear signal, or a less challenging condition; 2. On the other hand, if the number density is much smaller than one point per PSF area, the image turns into a collection of discrete point-like sources, causing both the nominator and the denominator of eq.(\ref{shear12PSF}) to become small differences of large numbers, which are well known sources for numerical errors. The second point simply means it is hard to measure the shapes of sources that are much smaller than the size of the PSF. In practice, we can avoid the second situation by increasing the observation/integration time.

\subsection{Testing the Treatment of the Pixelation Effect}
\label{test_pixelation}

When the scale radius of the PSF is less than $3$ or $4$ times the pixel size, the galaxy/noise images start to look pixelated, and the shear recovery accuracy may be strongly affected by the discrete nature of the CCD pixels. To reconstruct continuous images, we use 2D interpolation methods to insert finer grid points. The finer grid size is chosen to be $2^{m}$ ($m$ is an integer) times smaller than the original grid size, and at least less than a quarter of the PSF scale radius $R_{PSF}$. It is worth noting that interpolation of the pixelated image, if necessary, is always the first step in our shear measurement procedure.   

\begin{figure}
\centering
\includegraphics{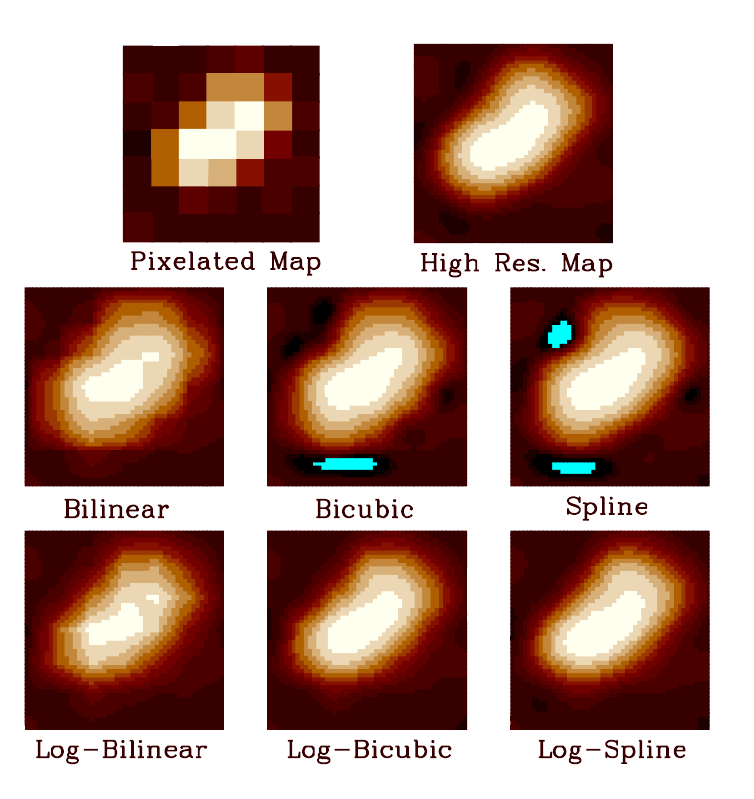}
\caption{The example of a pixelated image (upper-left), its high resolution counterpart (upper-right), and its interpolated images by the Bilinear, Bicubic, Spline, Log-Bilinear, Log-Bicubic, and Log-Spline methods (middle and lower panels). The blue regions have negative intensities. 
}
\label{maps_interpolation}
\end{figure}

An example of a pixelated image is shown in the up-left corner of fig.\ref{maps_interpolation}, for which the scale radius of the PSF is $0.5$. The high resolution images reconstructed by the three interpolation methods and their logarithmic extensions discussed in \S\ref{pixelation} are show in the lower half of fig.\ref{maps_interpolation}. The true high resolution image is on the up-right corner of the figure. By simply comparing the morphologies of the interpolated images by eye, one may already tend to conclude that the performances of the three conventional methods are improved if we interpolate the log of the data instead of the data itself. For instance, negative intensities (denoted as blue regions in the figure) are commonly found in the interpolated maps by the Bicubic and Spline methods, but absent in those by the Log-Bicubic or Log-Spline methods; the filamentary features produced by the Bilinear method become somewhat less prominent in the map processed by the Log-Bilinear method. 

\begin{figure}
\centering
\includegraphics{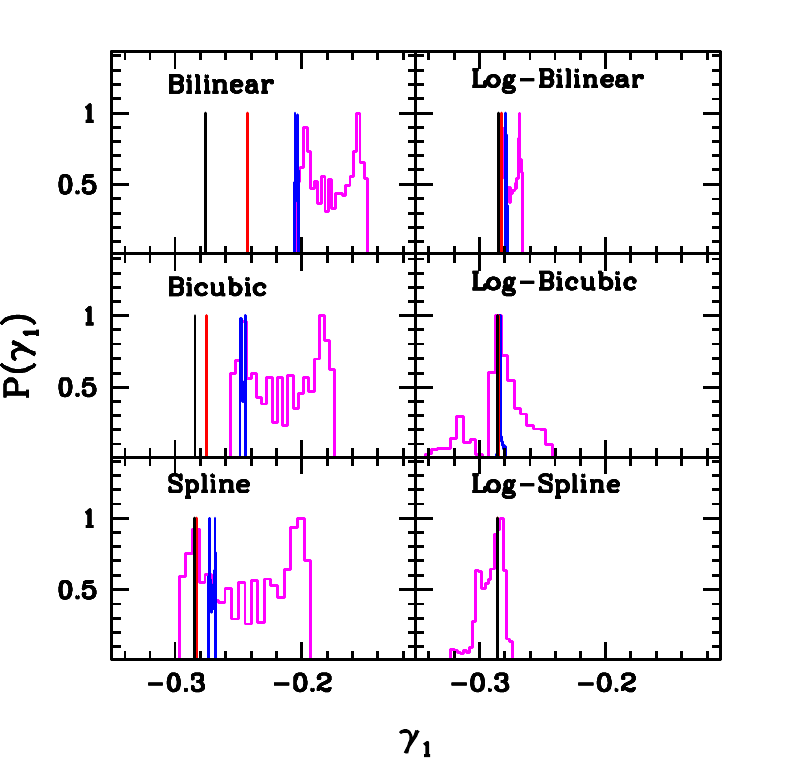}
\caption{The distribution of the shear values measured from the interpolated images of a {\it single} galaxy placed at random positions on the grid. The purple, blue, red, and black histograms correspond to the pixel-size-to-PSF-radius ratio of $2$, $1.5$, $1$, $0.5$ respectively. Each panel shows the results from a single interpolation method, whose name is indicated in the upper-left corner of the plot. All the histograms are normalized so that their peak values are one.  
}
\label{test_single_image}
\end{figure}

To test the interpolation methods more quantitatively, we may compare the shear values measured from the interpolated maps using eq.(\ref{shear12PSF}) (in the absence of noise). Clearly, all the interpolation methods should yield the same and correct shear estimates for a given galaxy and PSF if the PSF scale radius is much larger than the pixel size, \ie, in the absence of the pixelation effect. For small PSF sizes, as we have seen, the continuous images interpolated by different methods look unlike each other, resulting in possibly very different shear values. Moreover, since the source is sparsely sampled in this case, the original (pixelated) image, the interpolated image, and the measured shear values all depend on the relative positions of the pixels with respect to the source. In fig.\ref{test_single_image}, we show the distributions (as histograms) of the shear values estimated from a {\it single} galaxy that is placed at different/random locations on the grid. For simplicity and clarity, we do not include any photon noise here. The ratio of the galaxy size $R_G$ to the PSF scale radius $R_{PSF}$ is fixed at $5$. The ratio of the pixel size to $R_{PSF}$ is chosen to be $2$, $1.5$, $1$, $0.5$, the results of which are represented by the purple, blue, red, and black histograms respectively. The figure shows that as the pixel size decreases relative to the PSF size, the shear distributions converge more rapidly to a delta function at the correct position in methods with the prefix ``Log-'', manifesting again the value of the logarithmic extensions of the three classic interpolation methods.  

\begin{figure}
\centering
\includegraphics{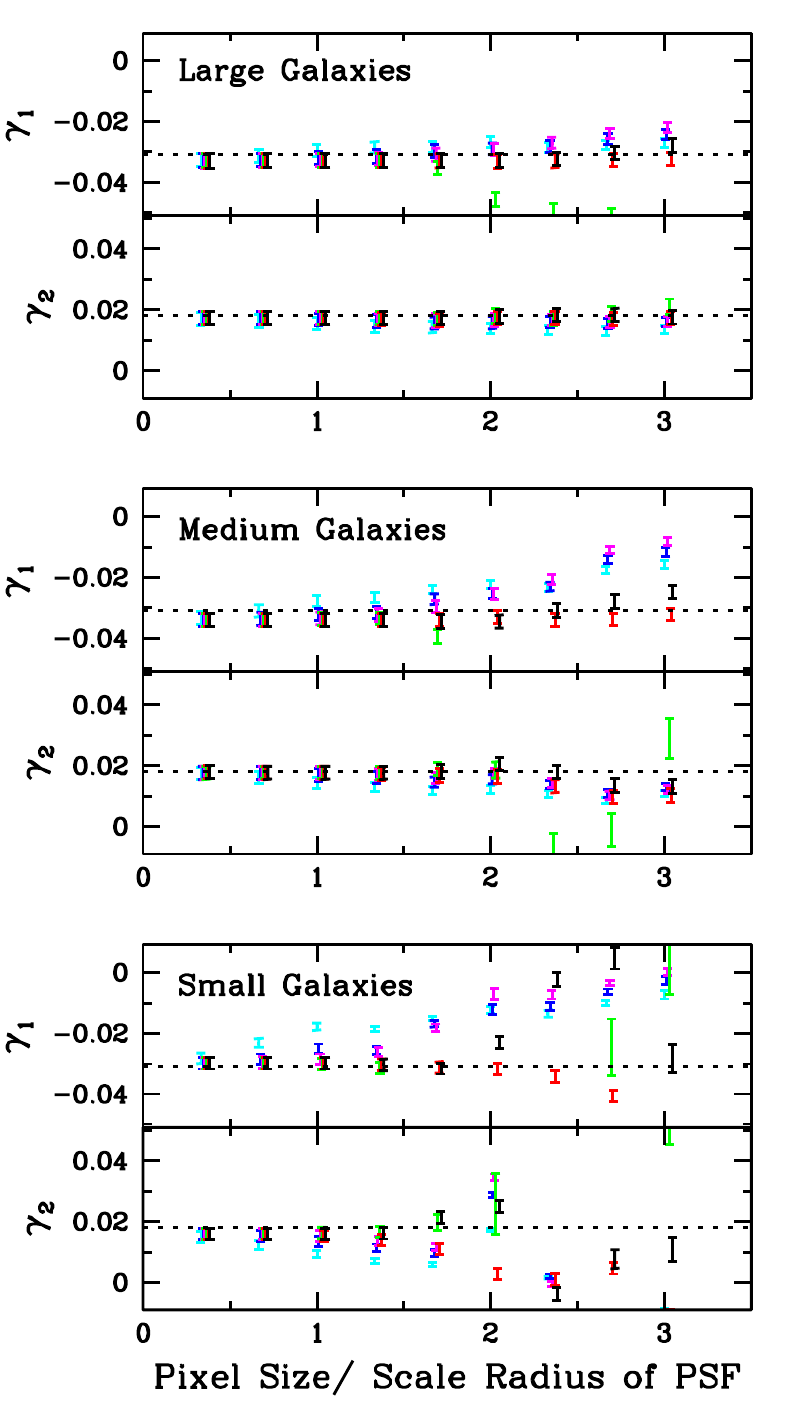}
\caption{The measured shear values plotted against the ratio of the pixel size to the PSF scale radius $R_{PSF}$. The upper, middle, and lower windows show the results from averaging over 10000 relatively large, medium, and small size galaxies respectively. The definition of the galaxy size in this example can be found in \S\ref{test_pixelation}. The cyan, blue, magenta, green, red, and black data points with $1\sigma$ error bars are the results from the Bilinear, Bicubic, Spline, Log-Bilinear, Log-Bicubic, and Log-Spline interpolation methods respectively. The input shear values are shown as dotted lines. 
}
\label{test_multi_image}
\end{figure}

Finally, let us find out which interpolation method is best suited to weak lensing. For this purpose, we test the accuracy of shear recovery with a large number of interpolated galaxy images. To make it a more convincing test, we use our morphologically rich irregular galaxies, each of which is generated by 1000 random steps. We consider three choices for the random walk step size and the galaxy scale radius $R_G$: ($0.5R_{PSF}$, $6.67R_{PSF}$), ($0.25R_{PSF}$, $3.33R_{PSF}$), and ($0.125R_{PSF}$, $1.67R_{PSF}$), referring to large, medium, and small galaxies respectively. The PSF we use is $W_B$. $\beta$ of the target isotropic Gaussian PSF is set to be $4/3$ of $R_{PSF}$. The cosmic shear ($\gamma_1$, $\gamma_2$) is chosen to be ($-0.031$, $0.018$). No photon noise is included. Our results are summarized in fig.\ref{test_multi_image}, in which we plot the measured shear values against the ratio of the pixel size to $R_{PSF}$. In the upper, middle, and lower panels of the figure, we report the results from averaging over 10000 large, medium, and small size galaxies respectively. The dotted lines refer to the input shear values. The cyan, blue, magenta, green, red, and black data points with $1\sigma$ error bars are from the Bilinear, Bicubic, Spline, Log-Bilinear, Log-Bicubic, and Log-Spline methods respectively. According to the figure, we can draw several conclusions:

1. As a sanity check, we confirm that all the interpolation methods work well when the PSF size is much larger than the pixel size;

2. Log-Bicubic and Log-Spline are the two most successful methods. Both of them can correctly recover the input shear as long as $R_{PSF}$ is about larger than a half of the pixel size, regardless of the galaxy size. Note that one should not expect any interpolation method to work well when $R_{PSF}\ll 0.5$ unless the source images are sufficiently smooth over the scale of the pixel size. 

3. The pixelation effect is less important for larger galaxies. For instance, by comparing the results in the three panels of fig.\ref{test_multi_image}, we see that the quality of shear recovery becomes increasingly poor for galaxies of smaller sizes for a given $R_{PSF}$. This is not surprising because the structures/shapes of large galaxies are better resolved than those of smaller ones. Meanwhile, it is encouraging to note that the Log-Bicubic and Log-Spline methods perform fairly well for $R_{PSF}\gsim 0.5$ even when the galaxy size is comparable to the PSF size.

\subsection{Testing the Overall Performance}
\label{overall}

\begin{figure}
\centering
\includegraphics{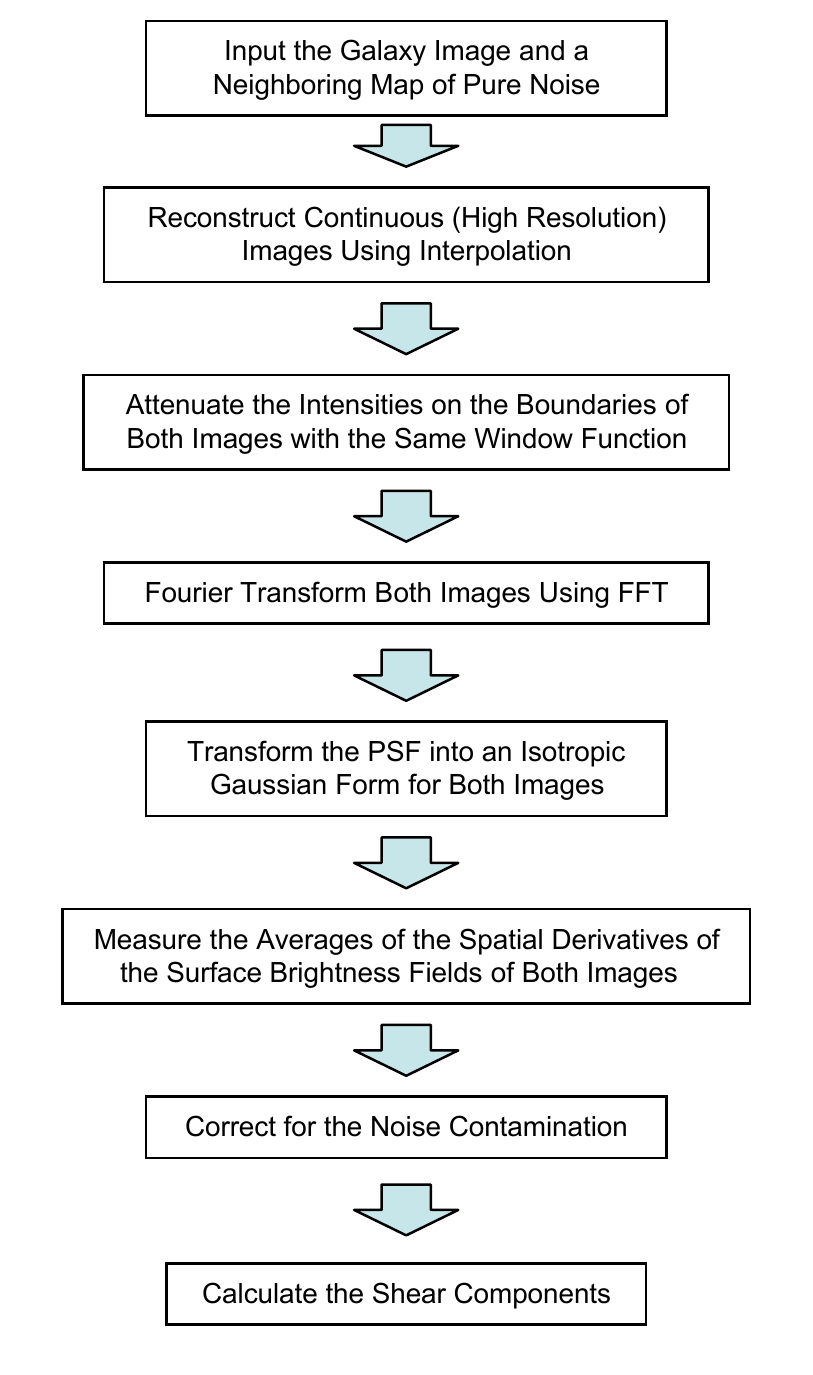}
\caption{The pipeline of our shear measurement procedures in the presence of both the photon noise and the pixelation effect. 
}
\label{pipeline}
\end{figure}

The purpose of this section is to test our shear measurement method under general conditions, \ie, in the presence of both photon noise and the pixelation effect. Fig.\ref{pipeline} shows the pipeline of the numerical procedures we take in general cases. A detailed explanation of each item in the graph has been given in \S\ref{systematics}. 

\begin{figure}
\centering
\includegraphics{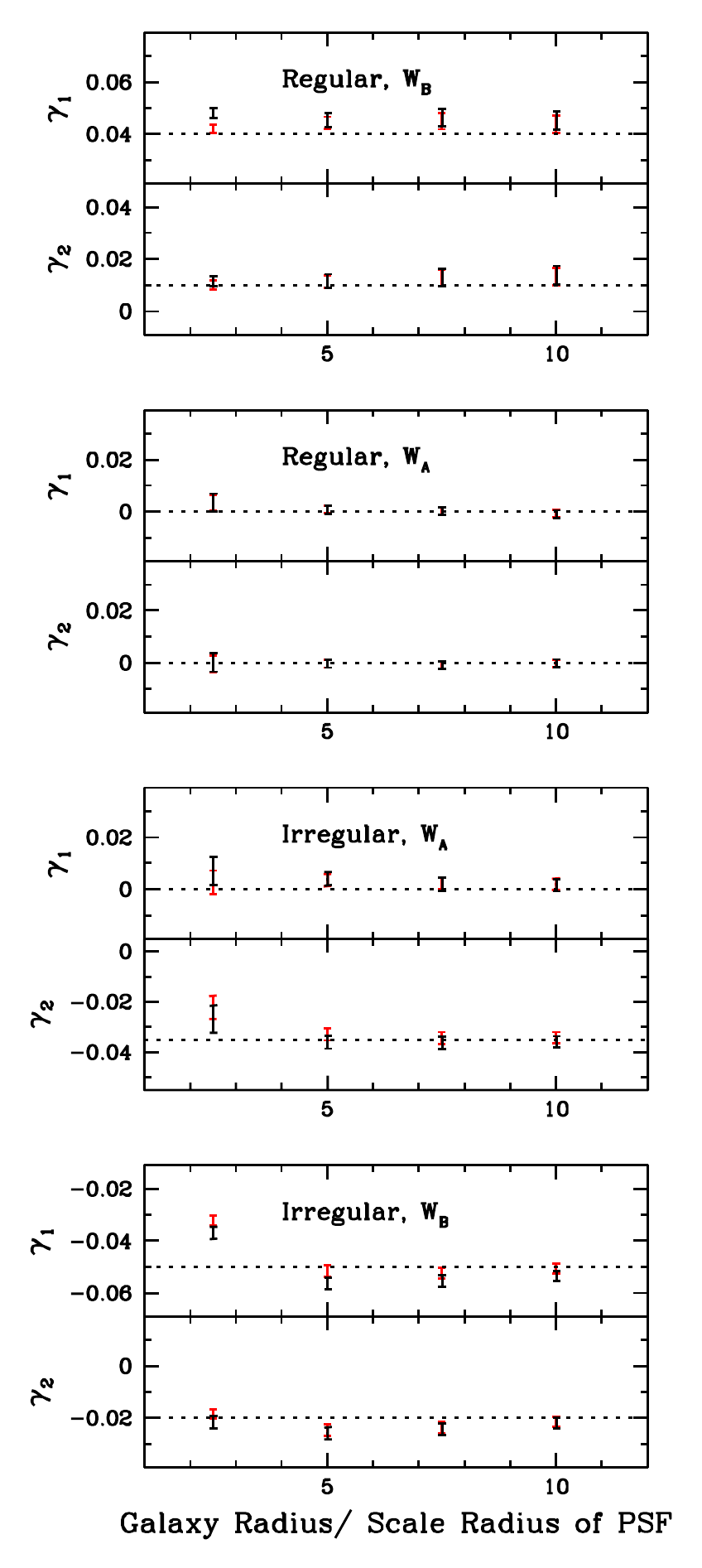}
\caption{The four panels show the accuracy of shear recovery with four different combinations of galaxy type (Regular/Irregular) and PSF form ($W_A$/$W_B$) as denoted at the top of each panel. For the results in each panel, we use 10000 mock galaxies, and scale them to four different sizes (galaxy radius equal to 2.5, 5, 7.5, 10 times the PSF scale radius) to recovery the input shear values. In all panels, the PSF scale radius is half of the pixel size. The red and black data points are from the Log-Bicubic and Log-Spline interpolation methods respectively. The input shear values are shown by dashed lines. From the top to the bottom panel, the ratio of the mean flux density of the noise to that of the galaxy ($I_N/I_G$) is 0.1, 0.5, 0.6, and 0.2 respectively.}
\label{overall_test}
\end{figure}

In fig.\ref{overall_test}, we show the shear recovery results for both regular and irregular galaxies with different PSF forms. In each panel, we use $10000$ mock galaxies and scale them to four different sizes (galaxy radius ranges from 2.5 to 10 times the scale radius of the PSF) to test the accuracy of shear recovery. In all panels, the PSF scale radius ($R_{PSF}$) is half of the pixel size, corresponding to roughly the maximum pixelation effect that can be treated by an interpolation method. The red and black data points are from the Log-Bicubic and Log-Spline methods respectively. The input shear values are shown by dashed lines. The scale radius of the target isotropic Gaussian PSF is always $1.5R_{PSF}$. To avoid aliasing powers in Fourier transformation, we use eq.(\ref{window_noise}) to filter the noise near the boundaries of the map. From the top to the bottom panel, the ratio of the mean flux density of the noise to that of the galaxy ($I_N/I_G$) is 0.1, 0.5, 0.6, and 0.2 respectively. 

The figure indicates that our method generally works well on galaxies of sizes that are at least a few times larger than the PSF size. Small discrepancies between the input shear values and the measured ones do exist when the galaxy size is comparable to the PSF size. The residual systematic errors will be studied with a much larger ensemble of galaxies in another work.

\section{Discussion and Conclusions}
\label{summary}

We have discussed how to correct for the systematic errors due to the photon noise and the pixelation effect in cosmic shear measurements. Our treatment of photon noise allows us to reliably remove the noise contamination to the cosmic shear even when the noise flux density is comparable with that of the sources. In principle, our method works regardless of the brightness of the noise, though when the noise is much brighter than the sources, one needs to worry about image selections. To deal with pixelated images, our approach is to reconstruct continuous images by interpolating the natural logarithms of the pixel readouts with either the Bicubic or Bicubic Spline method. This technique is accurate for the purpose of shear recovery as long as the scale radius of the PSF is larger than about a half of the pixel size, a condition which is almost always satisfied in practice. 

Despite the fact that our study has been based on the shear measurement method of Z08, a part of our methodology is generally useful for other shear measurement methods, or even other astronomical measurements as well. The most obvious thing to note is that the Log-Bicubic and Log-Spline interpolation methods are accurate image reconstruction approaches not only for weak lensing, but also for all kinds of other purposes. The way we remove the noise contamination from the shear signal can in principle also be considered in other shear measurements, in particular those that are based on measuring the multipole moments of the source images (\eg, \citealt{kaiser95} and its various extensions). 

\begin{figure}
\centering
\includegraphics{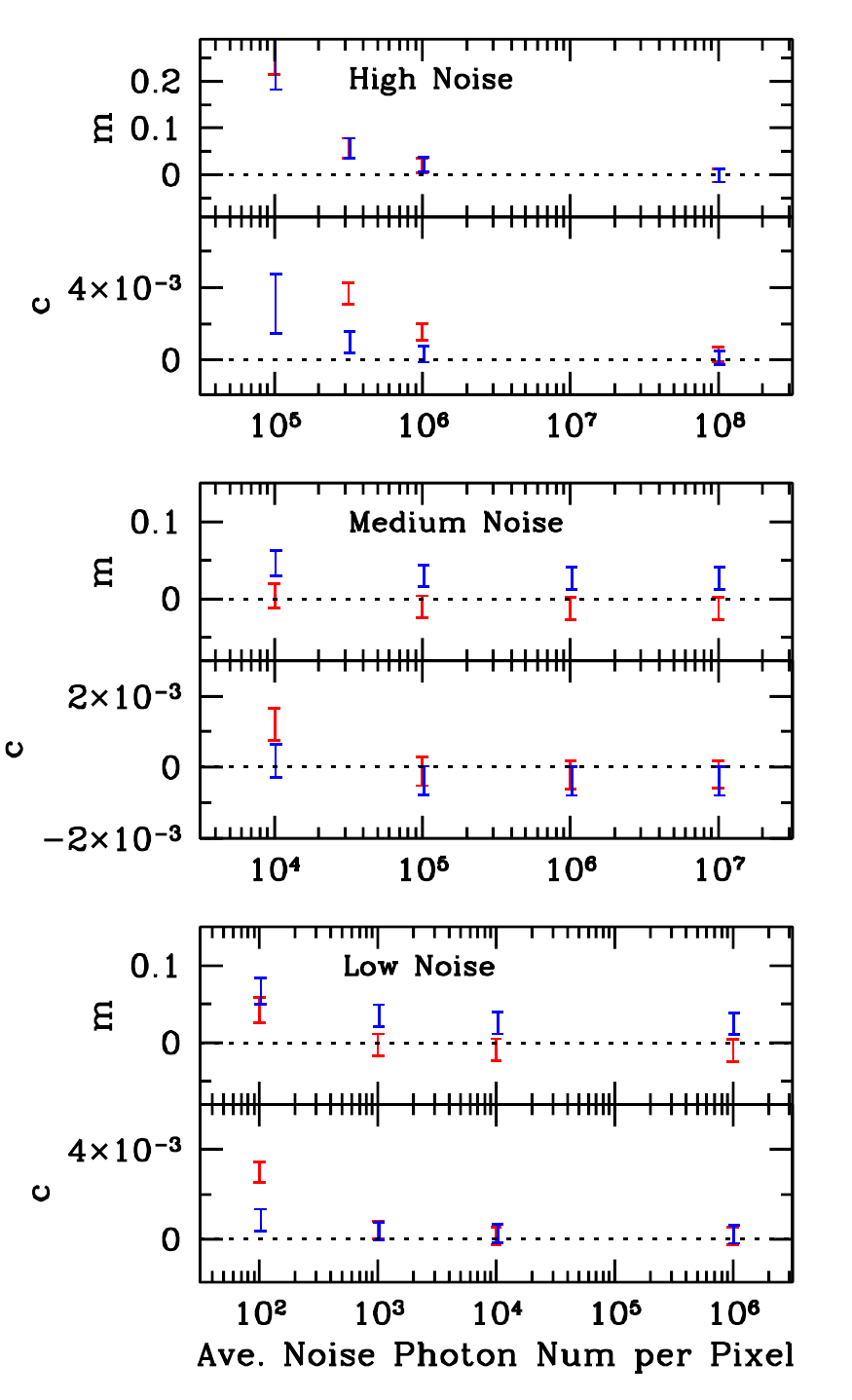}
\caption{We show that in the shear measurement method of Z08, how the shear recovery accuracy is affected by the exposure time (\ie, the mean number of background photons received per pixel). The results are expressed in terms of the multiplicative (m) and additive (c) bias parameters defined in eq.~(\ref{mc}). Each data point is achieved using five different sets of ($\gamma_1$, $\gamma_2$), \ie, (0.04, 0.04), (0.02, 0.02), (0, 0), (-0.02, -0.02), (-0.04, -0.04). The red and blue data points are for $\gamma_1$ and $\gamma_2$ respectively. For each input shear, 10000 mock galaxies are used to recover the shear. The same galaxies are used repeatedly for different exposure times so that the dependence of the shear recovery accuracy on the exposure time can be more clearly seen. The exposure time is denoted by the mean number of background photons received per pixel (x-axis). The three panels from the top to the bottom correspond to the signal-to-noise ratios (\ie, the ratio of the mean surface brightness of the source to that of the background noise) of 0.01, 0.1, 1 respectively, referring to the high, medium, and low noise cases as denoted at the top of the panels.
}
\label{Poisson}
\end{figure}

So far, our discussion has neglected the "photon counting" shot noise, which is always present due to the finite telescope exposure time. Indeed, it becomes the dominant source of photon noise for ground-based weak lensing survey because of the large sky background\footnote{Note that the mean of the background photon does not affect our shear measurement, because the method of Z08 uses the spatial derivatives of the surface brightness field. It is the fluctuation of the photon numbers across the pixels that may bias our shear estimator.}. Unlike the astronomical photon noise that we have discussed, the photon shot noise varies from pixel to pixel independent of the size of the PSF. Furthermore, the fluctuation amplitude of the shot noise is also dependent on the photon flux of the source. Therefore, it is hard to cleanly remove the contamination from the photon counting shot noise in our shear measurement. We do not intend to deal with this problem in this paper. Alternatively, we argue that the systematic shear measurement error due to the shot noise can be suppressed by simply increasing the telescope exposure time. More specifically, we argue that for any given tolerance level of the systematic error, there is a critical exposure time, beyond which the contamination from the shot noise is adequately suppressed.  To demonstrate this statement, in fig.\ref{Poisson}, we plot the accuracy of shear recovery under three different conditions: high, medium, and low noise level, which correspond to noise-to-source mean surface brightness ratio of 100, 10, and 1, or the signal-to-noise ratio of 0.01, 0.1, 1 respectively. The x axis in each plot is the mean number of background noise photons recorded on each pixel, which is proportional to the exposure time. The accuracy of shear recovery is expressed in terms of the multiplicative and additive bias parameters (commonly used by people in the weak lensing community), which are defined as:
\begin{equation}
\label{mc}
\gamma_{measured}=\gamma_{true}(1+m)+c
\end{equation}
For a perfect shear measurement, one should have $m=c=0$. To produce each ($m$, $c$) pair in fig.\ref{Poisson}, we use five different input shear values, \ie, (0.04, 0.04), (0.02, 0.02), (0, 0), (-0.02, -0.02), (-0.04, -0.04),  for ($\gamma_1$, $\gamma_2$) (which are reported using red and blue colors respectively), and 10000 mock irregular galaxies for each input ($\gamma_1$, $\gamma_2$). Note that to see the dependence of the shear recovery accuracy on the exposure time more clearly, we repeatedly use the same set of galaxies for different exposure times. In all cases, we set the galaxy radius to be 7.5 times the PSF radius. The later is set to be equal to the pixel size. We use Log-Spline as the interpolation method. The astronomical photon noise is not included, \ie, the photon counting noise is the only photon noise in this test.  According to the figure, the systematic errors due to the photon counting noise is clearly suppressed when exposure time is beyond some threshold. Not surprisingly, the low noise case requires the shortest exposure time to achieve the same shear recovery accuracy level. A more comprehensive test with a much larger ensemble of galaxies will be shown in a separate paper. 

The other sources of systematic errors that we have neglected include the high order corrections (\eg, $\gamma^2$, $\kappa^2$, $\gamma\kappa$) to our master equation [eq.(\ref{shear12PSF})], the spatial variations of the cosmic shear, etc.. These factors likely affect the measured shear values at percent levels on cosmic scales, which is important in the era of precision cosmology. For clustering lensing, the high order shear terms are more important because the shear is of order ten percent. This subject will be studied in a companion paper. 

This paper is a natural continuation of Z08 on the methodology of cosmic shear measurement. In another paper, we will further test this method with the data from the Shear TEsting Program (\citealt{heymans06,massey07}) and the GREAT08 program (\citealt{bridle09}), and also present results measured with real astronomical data. 

\section*{Acknowledgements}

JZ would like to thank Anthony Tyson and the anonymous referee for pointing out the importance of the photon counting shot noise, which was neglected in an earlier version of this paper. JZ is currently supported by the TCC Fellowship of Texas Cosmology Center of the University of Texas at Austin. JZ was previously supported by the TAC Fellowship of the Theoretical Astrophysics Center of UC Berkeley, where the majority of this work was done. 

\bibliographystyle{mn2e}

\vskip 1cm

\section*{Appendix -- Definitions of the Interpolation Methods}
\label{appendix}

In this appendix, we give the mathematical definitions of the three classic 2D interpolation methods: Bilinear, Bicubic, Spline. For their logarithmic extensions (\ie, Log-Bilinear, Log-Bicubic, Log-Spline), we only have one minor point to address at the end of this section.

The Bilinear method is the simplest of the three. Let us write the coordinates of the grid points as $(x_i,y_j)$ ($i,j=1,2,3...$), and the signals as $A(x_i,y_j)$. Suppose the point of our interest is $(x,y)$, which satisfies $x_i\le x\le x_{i+1}$ and $y_j\le y\le y_{j+1}$, the Bilinear method defines $A(x,y)$ in the following way:
\begin{eqnarray}
\label{A_bilinear}
A(x,y)&=&tuA(x_{i+1},y_{j+1})+(1-t)uA(x_i,y_{j+1})\\ \nonumber
&+&t(1-u)A(x_{i+1},y_j)+(1-t)(1-u)A(x_i,y_j)
\end{eqnarray}
where
\begin{eqnarray}
\label{tu}
&&t=(x-x_i)/(x_{i+1}-x_i)\\ \nonumber
&&u=(y-y_j)/(y_{j+1}-y_j).
\end{eqnarray}

The Bicubic method includes higher order terms of $t$ and $u$ to achieve smoothness of the interpolated function. It requires the user to specify not only the signal $A(x_i,y_j)$, but also the spatial derivatives $\partial A/\partial x$, $\partial A/\partial y$, and $\partial^2A/\partial x\partial y$ at every grid point $(x_i,y_j)$. Since the spatial derivatives of the signal are usually not known a priori, we estimate them using the finite-difference method:
\begin{eqnarray}
\label{dAdxdy}
\frac{\partial A}{\partial x}(x_i,y_j)&=&\frac{A(x_{i+1},y_j)-A(x_{i-1},y_j)}{x_{i+1}-x_{i-1}}\\ \nonumber
\frac{\partial A}{\partial y}(x_i,y_j)&=&\frac{A(x_i,y_{j+1})-A(x_i,y_{j-1})}{y_{j+1}-y_{j-1}}\\ \nonumber
\frac{\partial^2 A}{\partial x\partial y}(x_i,y_j)&=&\left[A(x_{i+1},y_{j+1})+A(x_{i-1},y_{j-1})\right.\\ \nonumber
&-&\left.A(x_{i+1},y_{j-1})-A(x_{i-1},y_{j+1})\right]\\ \nonumber
&/&\left[(x_{i+1}-x_{i-1})(y_{j+1}-y_{j-1})\right]
\end{eqnarray}
The interpolated function inside each grid square is written in the following polynomial form:
\begin{equation} 
\label{A_bicubic}
A(x,y)=\sum_{m=1}^{4}\sum_{n=1}^{4}c_{mn}t^{m-1}u^{n-1}
\end{equation}
The values of the sixteen parameters $c_{mn}$ are constrained using eq.(\ref{A_bicubic}) and the following three equations at the four corners of the grid square:
\begin{eqnarray}
\label{dAdxs}
&&\frac{\partial A}{\partial x}(x,y)=\sum_{m=2}^{4}\sum_{n=1}^{4}(m-1)c_{mn}t^{m-2}u^{n-1}\frac{dt}{dx}\\ \nonumber
&&\frac{\partial A}{\partial y}(x,y)=\sum_{m=1}^{4}\sum_{n=2}^{4}(n-1)c_{mn}t^{m-1}u^{n-2}\frac{du}{dy}\\ \nonumber
&&\frac{\partial^2 A}{\partial x\partial y}(x,y)=\sum_{m=2}^{4}\sum_{n=2}^{4}(m-1)(n-1)c_{mn}t^{m-2}u^{n-2}\frac{dt}{dx}\frac{du}{dy}\\ \nonumber
\end{eqnarray}
where $t$ and $u$ have been defined in eq.(\ref{tu}). 

The 2D Spline method simply refers to using the 1D Spline interpolation along each dimension. The 1D (cubic) Spline interpolation method works as follows: 

Given the values of a function $f(x)$ at a set of points $x_i$ ($i=1...N$), the form of the function in the interval between $x_j$ and $x_{j+1}$ is written as:
\begin{equation}
\label{A_spline}
f(x)=Hf(x_j)+Kf(x_{j+1})+Uf''(x_j)+Vf''(x_{j+1})
\end{equation} 
where 
\begin{eqnarray}
\label{HKUV}
&&H=\frac{x_{j+1}-x}{x_{j+1}-x_j}\\ \nonumber
&&K=1-H\\ \nonumber
&&U=\frac{1}{6}(H^3-H)(x_{j+1}-x_j)^2\\ \nonumber
&&V=\frac{1}{6}(K^3-K)(x_{j+1}-x_j)^2
\end{eqnarray}
and $f''(x_j)$ is the second derivative of the function $f$ at $x_j$. As a consistency check, one can easily show that $d^2f/dx^2=Hf''(x_j)+Kf''(x_{j+1})$. The value of the second derivatives are specified by requiring the first derivatives evaluated from the two sides of the grid point to be equal. Note that this requirement only provides $N-2$ equations, while there are $N$ second derivatives in total. The rest of the constraint comes from the boundary conditions on $f''(1)$ and $f''(N)$. In this paper, we simply set $f''(1)=f''(N)=0$, which yields the so-called {\it natural cubic spline}.  

Finally, we note that the ``Log'' based interpolation methods are all well defined except when the readouts of some pixels are zero. This is a very rare case in practice due to the presence of noise. However, this situation can in principle exist in simulations. To cure this problem, one can either change the zeros into tiny positive numbers, or simply avoid interpolating the regions with zeros. The second option says that if a grid square (regarding the Log-Bilinear and Log-Bicubic methods) or a unitary segment (regarding the Log-Spline method) contains any zero readouts in their four corners or two ends, the finer grid points within them are all set to have zero values. The rest of the grid squares/segments are interpolated independently as usual. Note that in the Log-Spline method, this means the Spline interpolations are carried out only in those nonzero segments that are isolated by the zeros. These two choices usually work similarly well. However, when there are extended regions of zero readouts, we find that the second option is better, because it avoids introducing artificial high order fluctuations in the zero regions by methods like Log-Bicubic or Log-Spline.

\label{lastpage}

\end{document}